\documentclass[usenatbib,usegraphicx,usedcolumn]{mn2e}
\usepackage{times,amssymb,amsmath}
\title[AGN evolution and feedback]{Cosmological growth and feedback from supermassive black holes}
\author[P. Mocz et. al.]{P. Mocz$^{1,2}$\thanks{E-mail: pmocz@cfa.harvard.edu (PM); acf@ast.cam.ac.uk (ACF); kmb@astro.ox.ac.uk (KMB)}, A.C. Fabian$^{2}$\footnotemark[1] and Katherine M. Blundell$^{3}$\footnotemark[1] \\
$^{1}$Harvard University, Cambridge, MA 02138, USA\\
$^{2}$Institute of Astronomy, Madingley Road, Cambridge CB3 0HA, UK\\
$^{3}$Astrophysics, University of Oxford, Keble Road, Oxford OX1 3RH, UK}
\begin{document}

\date{subm. to MNRAS, 29 September 2012, accepted 19 April 2013}

\pagerange{\pageref{firstpage}--\pageref{lastpage}} \pubyear{2013}

\maketitle

\label{firstpage}

\begin{abstract}
We develop a simple evolutionary scenario for the growth of supermassive black holes (BHs), assuming growth due to accretion only, to learn about the evolution of the BH mass function from $z=3$ to $0$ and from it calculate the energy budgets of different modes of feedback. We tune the parameters of the model by matching the derived X-ray luminosity function (XLF) with the observed XLF of active galactic nuclei. We then calculate the amount of comoving kinetic and bolometric feedback as a function of redshift, derive a kinetic luminosity function and estimate the amount of kinetic feedback and $PdV$ work done by classical double Fanaroff-Riley II (FR~II) radio sources. We also derive the radio luminosity function for FR~IIs from our synthesized population and set constraints on jet duty cycles. Around $1/6$ of the jet power from FR~II sources goes into $PdV$ work done in the expanding lobes during the time the jet is on. Anti hierarchical growth of BHs is seen in our model due to addition of an amount of mass being accreted on to all BHs independent of the BH mass. The contribution to the total kinetic feedback by active galaxies in a low accretion, kinetically efficient mode is found to be the most significant at $z<1.5$. FR~II feedback is found to be a significant mode of feedback above redshifts $z\sim 1.5$, which has not been highlighted by previous studies. 
\end{abstract}


\section{Introduction}\label{sec:intro}

As supermassive black holes (BHs) at centres of galaxies grow through mass accretion, they also exhibit different types of feedback in radiative and kinetic forms. Understanding the energy budget of the various feedback components is important because they may couple to the surrounding environment in distinct ways. Most, if not all, galaxies are believed to host supermassive BHs \citep{1998AJ....115.2285M}, and their roles in galaxy evolution cannot be neglected. Mechanical feedback from active galactic nuclei (AGN) has recently been invoked to be an important part of galaxy formation and the heating of inflowing gas, providing energy to suppress star formation (first steps have been taken by a number of groups: \citealt{2000MNRAS.311..576K,2006MNRAS.365...11C,2006ApJS..166....1H,2006MNRAS.370..645B}). Without including such an additional heating mechanism, cosmological models overpredict the number of faint and bright galaxies (e.g. \citealt{1978MNRAS.183..341W,1991ApJ...379...52W,2003ApJ...599...38B}) and predict that the largest galaxies in the present epoch are blue and star forming rather than `red and dead', contrary to observations. Additionally, BH feedback is known to play an important role in energy input in galaxy clusters (e.g. \citealt{2003MNRAS.344L..43F}) and galaxy groups (e.g. \citealt{2010MNRAS.406..822M}). Accounting for different feedback modes more carefully may augment our picture of galaxy evolution.

It can be difficult to estimate the cosmic history of the energy budget of certain feedback modes directly, though recent calculations for mechanical feedback from AGN, relying on various assumptions, have been carried out by \cite{2008MNRAS.388.1011M}, \cite{2008MNRAS.383..277K}, and \cite{2009MNRAS.395..518C}. One of the more challenging modes to calculate, for example, is mechanical feedback of classical double Fanaroff-Riley II (FR~II) sources. Here one must carefully treat the complicated physics of radio emission and particle acceleration in the hotspots. The lobes are radio-luminous because of the synchrotron radiation from ultrarelativistic electrons accelerated at the hotspots, but this may account for only a small fraction of the total energy output, the rest of which may be used to do $PdV$ work in expanding the lobes or be stored as internal energy. Additionally, the duty cycle of jets is not well known. A further complicating factor is the `youth-redshift' degeneracy \citep{1999Natur.399..330B}, namely that older radio sources do not survive above previous radio survey flux limits. Therefore, in this work we seek an approach to estimate the cosmic volume-averaged energies from different modes of feedback which is suited to obtain a better handle on FR~II feedback than approaches taken in previous studies. In our approach we do not make assumptions about jet duty cycle or relations between the total radio and X-ray luminosity of a source. We also do not use the observed radio luminosity function (RLF) of radio galaxies directly to make our estimates due to the confounding effects of jet duty cycle and the `youth-redshift' degeneracy.

In our approach, we first reconstruct the cosmic history of accretion on to supermassive BHs using a simple volume-averaged evolutionary scenario/prescription for accretion. Our approach is quite different from previous work because we assume directly a physically-motivated functional form for the volume-averaged accretion rate $\langle \dot{M}\rangle$ as a function of BH mass and cosmic time, with a limited number of free parameters to tune that affect the evolution in independent ways. We do expect that most galaxies were more active in the past, during the quasar era, $1.5<z<3$, where most radio activity is seen \citep{2000MNRAS.319..121J,2001AJ....122.2177B}. We use observational constraints to tune the parameters: namely, the local BH mass function (BHMF) which has been estimated using well-known correlations between the BH mass, bulge luminosity, and stellar velocity dispersion of the galaxy (e.g. \citealt{2004MNRAS.351..169M}) and the AGN luminosity function (LF) in the $2$-$10$~keV luminosity range of  $10^{41.5}$--$10^{46.5}$~erg~s$^{-1}$, which has been characterized up to redshift $z=3$ \citep{2003ApJ...598..886U}.

In our evolutionary description, we assume that from $z=3$ to $0$ BH growth has occurred due to accretion only. In general, cosmic BH growth can be due to either BH-BH merger events or accretion (either secular or triggered by galaxy mergers) and the dominant process as a function of redshift are not fully understood. However, observational evidence suggests that much of AGN evolution is secular from $z=2$ to the present epoch \citep{2011MNRAS.417.2721O,2012ApJ...744..148K}. BH growth due to accretion only is also enough to explain the local BH mass density (\citealt{1999MNRAS.303L..34F,2002ApJ...565L..75E,2002MNRAS.335..965Y}, see also \citealt{2004A&A...420L..23K}). Such arguments imply that most BHs have an accretion efficiency of $\eta\sim 0.1$. BH growth due to mergers is only expected to be dominant for high mass BHs (with massive dark haloes) \citep{2002MNRAS.333..353C}. Therefore, it is plausible that BH growth due to merging events is negligible for $z\leq 3$. With this assumption, the BHMF can easily be integrated backwards in time using a continuity equation \citep{1992MNRAS.259..725S} with our prescription for $\langle \dot{M}\rangle$ to obtain the history of the evolution of BHs in the centres of galaxies. However, a fully detailed analysis would include the significance of mergers, which are a part of many theoretical models for quasar evolution for triggering gas infall and could potentially be responsible for triggering the quasar peak at $z=2$--$3$.

To link accretion with feedback, we use a physically-based model for AGN based on accretion in microquasars to explain the ratio of kinetic and radiative released energy as a function of BH accretion rate (discussed in, for example, \citealt{2008MNRAS.388.1011M}). A BH may accrete in three different modes, depending on its accretion rate: a low accretion, kinetic (LK) mode, a high accretion radiative (HR) mode (no jets), and a high accretion kinetic (HK) mode (powerful radio jets). After reconstructing the cosmic evolution from our prescription and observational constraints of the BHMF and X-ray luminosity function (XLF), we can construct an RLF to determine the duty cycle of powerful jets. 

There have been a number of studies that investigate the evolutionary growth of BHs. Many trace the evolution of BHs using the LF of AGN and integrate the continuity equation to obtain a local BHMF which is matched with the observationally derived one. Alternatively, the continuity equation has been integrated backwards in time to obtain an evolving BHMF. Some studies assume fixed Eddington ratios and accretion efficiencies for populations or sub-populations \citep{2004MNRAS.351..169M,2009MNRAS.396.1217R}, some assume redshift or mass-dependent Eddington ratios \citep{2009ApJ...690...20S}, some assume a so-called `Fundamental Plane' relation between the intrinsic (unbeamed) radio luminosity of the jet core, accretion-powered X-ray luminosity, and BH mass at all redshifts \citep{2004MNRAS.353.1035M,2008MNRAS.388.1011M}, and others assume a log-normal or power-law distribution of Eddington ratios \citep{2008MNRAS.390..561C,2010ApJ...725..388C}. From these studies, an anti hierarchical growth of BHs has been found (meaning that most lower mass BHs formed later in cosmic time than higher mass BHs), with possible reversal of the downsizing at $z\simeq 2$ and higher redshifts \citep{2008MNRAS.388.1011M}. Downsizing was first observationally found using X-ray surveys in 2003 \citep{2003ApJ...598..886U}.

In this paper, we take a different approach by considering a simple evolutionary scheme for accretion on to BHs. We use the model to derive the evolving BHMF from the local one using the continuity equation, and attempt to reconstruct the observed XLF to learn what sort of an evolutionary scenario may be possible and thereby infer a plausible reason for the anti hierarchical growth. The real evolutionary mechanism of a single BH is, of course, more complex, but we can gain insight into volume-averaged average BH growth as a function of cosmic time and what may be responsible for the observed anti hierarchical growth. We then calculate the volume-averaged kinetic feedback from jets as a function of redshift. We also construct a kinetic luminosity function (KLF) as well as an RLF for FR~II sources with analytic model developed for FR~II sources in \cite{2011MNRAS.413.1107M} and estimate jet duty cycle as a function of redshift.

In \S~\ref{sec:mbha} we describe a physically motivated model for the relation between kinetically and radiatively released power and the accretion rate of BHs. In \S~\ref{sec:evo} we describe our simple framework for the evolution of BHs. In \S~\ref{sec:results} we present the results of matching the observed XLF with our model as well as the derived population properties and calculation of different types of feedback. Finally in \S~\ref{sec:disc} we discuss the implications of our results. We adopt standard cosmological parameters $\Omega_{\rm M}=0.3$, $\Omega_\Lambda=0.7$ and $H_0=70~$km$~$s$^{-1}~$Mpc$^{-1}$ in this work.

\section{Supermassive black hole accretion}\label{sec:mbha}

The accretion rate of a BH may be described in Eddington units as $\dot{m}\equiv \eta \dot{M} c^2/L_{\rm{Edd}}$, where $M$ is the BH mass, $\eta$ is the accretion efficiency and $L_{\rm{Edd}}$ is the Eddington luminosity, $L_{\rm{Edd}} = 4\pi G M m_{\rm{p}} c / \sigma_{\rm T}$ where $m_{\rm{p}}$ is proton mass and $\sigma_{\rm T}$ is the Thomson cross-section. The accretion efficiency, $\eta$, depends on the inner boundary conditions of the accretion flow, and in the classical no-torque case has a value of $0.057$ for a Schwarzschild BH (no spin) up to a value of $0.42$ for a maximally spinning Kerr BH \citep{1973blho.conf..343N,2008MNRAS.388.1011M}. The accretion efficiency tells us the maximal amount of potential energy per unit time that can be extracted from the BH. The quantity $\dot{M}$ is the rate of mass accretion on to the BH. 

This power output of an accreting BH may be in radiative or kinetic form, and the efficiencies of these processes depend on both $\eta$ and the nature of the accretion flow. The source has bolometric luminosity $L_{\rm{bol}}$, from which the Eddington ratio can be defined: $\lambda \equiv L_{\rm{bol}} / L_{\rm{Edd}}$. The radiative efficiency of the source is: $\epsilon_{\rm{rad}} \equiv L_{\rm{bol}} / (\dot{M}c^2)$ (so that $\lambda = \epsilon_{\rm{rad}} \dot{M}c^2 / L_{\rm{Edd}} $). In addition, BHs may output kinetic power with efficiency $\epsilon_{\rm{kin}} \equiv  L_{\rm{kin}} / (\dot{M}c^2)$. If the kinetic power output is primarily in the form of double jets, then the single jet power is $Q_{\rm jet}=L_{\rm kin}/2$. The kinetic output may also be in the form of winds \citep{2007ApJ...668L.103B}.

An accreting BH may grow in different modes, with different ratios of kinetic and radiative power outputs. Here we will use a simple physically motivated model for modes of BH growth (see \citealt{2008MNRAS.388.1011M}, \citealt{2008MNRAS.383..277K} and references therein). The picture comes from an analogy of supermassive BHs with their less massive accreting counterparts: microquasars. We can describe the kinetic and radiative output as a function of the accretion rate $\dot{m}$. Above a critical accretion rate, $\dot{m}_{\rm crit}$, a BH is radiatively efficient, with $\lambda\propto \dot{m}$. We adopt a value of $\dot{m}_{\rm crit} = 3\times 10^{-2}$ (as in \citealt{2008MNRAS.388.1011M}), at which $\lambda = \lambda_{\rm crit} \simeq \dot{m}_{\rm crit}$. A BH accreting above the critical rate may be in one of two different physical states, a high accretion, kinetic (HK) state, i.e., one with powerful radio jets, or a high accretion, radiative (HR) state (no jets). The fraction of sources with jets switched on we will characterize by a jet duty cycle $f$, which can be a function of cosmic time. Below $\dot{m}_{\rm crit}$, a BH will evolve in a kinetically efficient, radiatively inefficient mode with $\lambda\propto \dot{m}^2$. This mode is the low accretion, kinetic (LK) mode.

We define the kinetic luminosity to be related to the Eddington luminosity by the factor $\lambda_{\rm kin}(\dot{m})$, that is: $L_{\rm{kin}} = \lambda_{\rm kin}(\dot{m}) L_{\rm{Edd}}$. In the HK mode $\lambda_{\rm kin}(\dot{m})\propto \dot{m}$ and in the HR mode $\lambda_{\rm kin}(\dot{m}) = \dot{m}_{\rm crit}$. We use the following parametrizations for $\lambda(\dot{m})$ and $\lambda_{\rm kin}(\dot{m})$, so that the parametrizations are differentiable and easy to invert:
\begin{equation}
\lambda(\dot{m}) = \left( \left( \frac{\dot{m} }{\dot{m}_{\rm crit}}\right)^{-2}  +  \left(  \frac{\dot{m} }{\dot{m}_{\rm crit}} \right)^{-1} \right)^{-1}
\label{eqn:p1}
\end{equation}
\begin{equation}
\lambda_{\rm kin}(\dot{m}) = 
\begin{cases}
\dot{m}  & \text{~HK~mode~above~$\dot{m}_{\rm crit}$}\\
\dot{m}_{\rm crit}
\left( \left( \frac{\dot{m} }{\dot{m}_{\rm crit}} \right)^{-1}  +  1\right)^{-1}  & \text{~HR~mode~above~$\dot{m}_{\rm crit}$}\\
\end{cases}\label{eqn:p2}
\end{equation}

\begin{figure}
\centering
\includegraphics[width=0.47\textwidth]{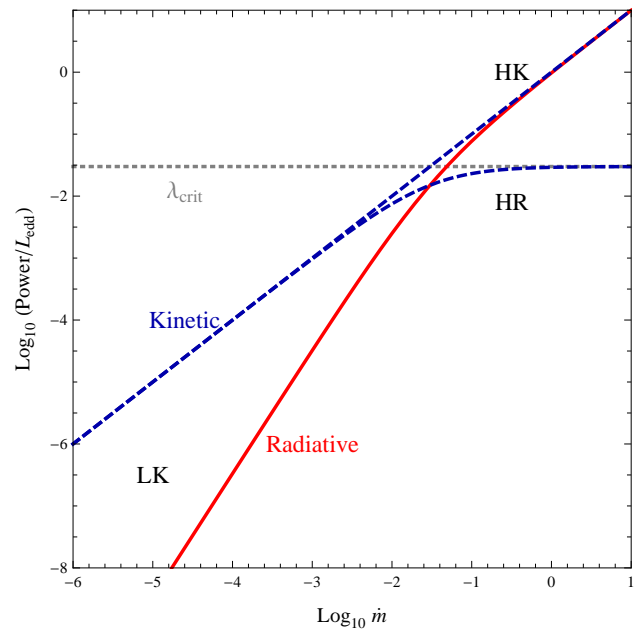}
\caption{Kinetic and radiative released power in Eddington units as a function of BH accretion rate, $\dot{m}$, assuming the parametrizations of equations (\ref{eqn:p1}) and (\ref{eqn:p2}). The horizontal dotted line is at $\lambda_{\rm crit}$. At low accretion rates, sources accrete in a radiatively inefficient (LK) mode. At high accretion rates, sources may accrete in either a kinetically efficient mode (HK, i.e., a radio-loud source with jets) or kinetically inefficient mode (HR, i.e., a radio-quiet source with weak/no jets).}
\label{fig:scheme}
\end{figure}

The relationship between accretion rate and released power in kinetic and radiative form is presented in Figure~\ref{fig:scheme}.
In this picture, powerful radio jets are episodic events \citep{2005MNRAS.361..633N,2011MNRAS.412..705B} that may happen when a source is accreting above the critical rate $\dot{m}_{\rm crit}$ with some duty cycle $f$. Sources with accretion rates below the critical level are assumed to have a high amount of kinetic output (compared to radiative output) all the time. It may be a reasonable assumption that powerful jetted sources behave differently than weaker jetted ones (which we assume are more stable and are on all the time) as the evolution of the comoving space density of powerful radio sources is known to be different from that of low luminosity radio sources \citep{2009ApJ...696...24S,2011MNRAS.413.1054M}. Also, the low luminosity Fanaroff-Riley I (FR~I) sources, the less-powerful counterparts of FR~IIs, are known to reside in denser environments than FR~IIs out to $z\sim 0.5$ \citep{2008AJ....135.1311A}, which means that FR~Is reside in environments with higher particle density and may have a more constant fuel supply. In addition, at the centres of X-ray bright cool core clusters an FR~I source is almost always found \citep{1990AJ.....99...14B,2009ApJ...704.1586S}, indicating that less-powerful jet activity may be more stable and long-term processes.

\section{The evolution of supermassive black holes}\label{sec:evo}

Here we develop a simple physically and observationally motivated model for the evolution of supermassive BHs in a volume averaged sense. We assume that the growth and energy output of the BHs are due to only mass accretion from an initial cosmic time corresponding to an initial redshift, $z_{\rm i}$ up to the present, $z=0$. We seek to reconstruct the BHMF, $M\frac{\partial N(M,t)}{\partial M}$, where $M$ is the BH mass and $N(M,t)$ is the number of BHs with mass $M$ per unit comoving volume as a function of time (or redshift). Integrating $M\frac{\partial N(M,t)}{\partial M}$ over some mass range at a fixed time gives the total mass of BHs within the mass range per unit comoving volume. Performing the integral over $\frac{\partial N(M,t)}{\partial M}$ gives the number density of BHs. Assuming that the total number of BHs is constant with time and that growth happens only due to accretion, the continuity equation \citep{1992MNRAS.259..725S} describes the evolution of the mass function:
\begin{equation}
\frac{\partial }{\partial t} \frac{\partial N(M,t)}{\partial M} +\frac{\partial}{\partial M} \left[ \frac{\partial N(M,t)}{\partial M} \langle\dot{M}(M,t) \rangle  \right] = 0
\label{continuityEqn}
\end{equation}
where $\langle\dot{M}(M,t) \rangle$ is the mean accretion rate for all BHs of mass $M$ at cosmic time $t$. The right-hand side of the equation would be equal to a source function if one wanted to describe merger events or the formation of new BH seeds.

Here, we consider a simple scheme for the evolution of BHs. The unitless accretion rate, $\dot{m}$, is proportional to $\dot{M}/M$. 
As a BH grows larger, $M$ increases which decreases $\dot{m}$ due to the $1/M$ proportionality. $\dot{M}$ is also expected to decrease with time, at least in an average sense, due to the depletion of available infalling matter, reduction of infalling material by AGN feedback, or perhaps some other physical mechanism that shuts down accretion (of course, an individual BH may have $\dot{M}$ increase with time during some periods in its evolution, for example a change in spin may increase the accretion rate). We assume that at an initial redshift $z_{\rm i}$, corresponding to initial lookback time $t_{\rm i}$, all the BHs are accreting brightly, with $\langle \dot{m} \rangle \geq 1$, and accretion rates will fall exponentially with time. In our evolutionary scenario, accretion near the Eddington limit and super-Eddington accretion are important at $z\simeq z_{\rm i}$, however $z_{\rm i}$ is chosen large enough so that super-Eddington accretion is not at all prevalent by the redshift $z=3$ and older Universe, which is where we have observational constraints and will be testing our model. What happens in the model for $z>3$ is not an important consideration for our results and is just an extrapolation of our model for $z<3$.

We assume that at time $t_{\rm i}$ a BH of initial mass $M_{\rm i}$ accretes matter, on average, at rate
\begin{equation}
\langle \dot{M}(M_{\rm i}) \rangle  = \frac{1}{\eta} \frac{4\pi G m_{\rm{p}} }{c \sigma_{\rm T}}  (M_{\rm i}+ M_{\rm{const}} )
\label{eqn:scenario}
\end{equation}
where $M_{\rm{const}}$ corresponds to a constant amount of mass being accreted by BHs of all masses in a time interval $\eta \frac{c \sigma_{\rm T}}{4\pi G m_{\rm{p}} } $. The variable $M_{\rm{const}}$ is a free parameter in our model. If $M_{\rm{const}} = 0$ then equation~(\ref{eqn:scenario}) corresponds to all BHs accreting with a rate $\langle\dot{m} \rangle = 1$. The presence of the $M_{\rm{const}}$ term allows low-mass and high-mass BHs to potentially have different accretion histories in the evolution, which is motivated by observations that suggest anti hierarchical growth of BHs.

Next we assume that the amount of mass available to be accreted by a BH may decline with time and therefore add exponential decay factors to the time evolution. That is, 
\begin{equation}
\langle \dot{M}(M_{\rm i},t) \rangle = \frac{1}{\eta} \frac{4\pi G m_{\rm{p}} }{c \sigma_{\rm T}}  (M_{\rm i} \rmn{e}^{\frac{t-t_{\rm i}}{\tau_1}} + M_{\rm{const}}\rmn{e}^{\frac{t-t_{\rm i}}{\tau_2}} )
\label{eqn:nM}
\end{equation}
where $\tau_1$ and $\tau_2$ are free parameters (note that $t$ here is the lookback time; $t=0$ corresponds to $z=0$).

We wish to find $\langle\dot{M}(M,t) \rangle$, the average accretion rate as a function of lookback time and mass, from this scheme. We assume that all BHs at the present epoch (and at $z<z_{\rm i}$) of a given mass $M$ had similar long-time-scale accretion histories so that they all originated from BHs of similar initial mass $M_{\rm i}$ at $t_{\rm i}$. That is, for a given $M$ and $t$ we solve
\begin{equation}
M = M_{\rm i} + \int_t^{t_{\rm i}} \langle \dot{M}(M_{\rm i},t^\prime) \rangle \ dt^\prime
\end{equation}
for $M_{\rm i} = M_{\rm i}(M,t)$, which yields
\begin{equation}
M_{\rm i}(M,t) = \frac{M + \frac{1}{\eta}\frac{4\pi G m_{\rm{p}} }{c \sigma_{\rm T}}M_{\rm{const}}\tau_2\left( \rmn{e}^{\frac{t-t_{\rm i}}{\tau_2}}-1  \right) }
{1 - \frac{1}{\eta}\frac{4\pi G m_{\rm{p}} }{c \sigma_{\rm T}}\tau_1 \left( \rmn{e}^{\frac{t-t_{\rm i}}{\tau_1}}-1  \right) }.
\label{eqn:Mi}
\end{equation}
Thus the average accretion rate as a function of time and mass, needed for the continuity equation, is found:
\begin{equation}
\langle\dot{M}(M,t) \rangle = \langle \dot{M}(M_{\rm i}(M,t),t) \rangle
\label{eqn:avgM}
\end{equation}
where $M_{\rm i}(M,t)$ comes from equation~(\ref{eqn:Mi}) and is substituted into equation~(\ref{eqn:nM}). The accretion efficiency $\eta$ is a free parameter in the model.

Knowing $\langle\dot{M}(M,t) \rangle$ and a boundary condition allows us to numerically solve for the evolution of $\frac{\partial N(M,t)}{\partial M}$ using the continuity equation (equation~(\ref{continuityEqn})). Equation~(\ref{eqn:avgM}) tells us $\langle\dot{M}(M,t) \rangle$. As for the boundary condition, we use the $10^6$--$10^{10}~M_\odot$ local BHMF derived by \cite{2004MNRAS.351..169M}. We set $z_{\rm i}=6$, but will solve the BHMF from $z=0$ up to $3$, where the XLF is constrained \citep{2003ApJ...598..886U}. We derive the evolving BHMF for masses $10^6$--$10^{10}~M_\odot$. The local BHMF is extrapolated for masses beyond $10^{10}~M_\odot$ in order to do so. An advantage of integrating the continuity equation backwards is that we will not have to make assumptions about the number density of BHs below $10^6~M_\odot$.

\subsection{Deriving the X-ray luminosity function}\label{sec:xlf}

In order to learn whether the simple evolution scheme we described can explain observations and find possible values for the free parameters $\eta$, $\tau_1$, $\tau_2$, and $M_{\rm{const}}$, we calculate the predicted hard XLF as a function of $z$ and compare it to the luminosity-dependent density evolution (LDDE) hard XLF derived from observation in the study by \cite{2003ApJ...598..886U}. A number of subsequent studies have also characterized the XLF, including \cite{2005ApJ...635..864L}, \cite{2010MNRAS.401.2531A}, and \cite{2011PASJ...63S.937U}. Using the XLF \cite{2003ApJ...598..886U} allows one to more directly compare our results to \cite{2008MNRAS.388.1011M}, which uses the same XLF as a constraint.

The LDDE XLF of \cite{2003ApJ...598..886U} is parametrized as follows:
\begin{equation}
\frac{\rmn{d} \phi (L_{\rm X},z)}{\rmn{d} \log L_{\rm X}} = A \left[  \left( \frac{L_{\rm X}}{L_*}\right)^{\gamma_1} + \left( \frac{L_{\rm X}}{L_*}\right)^{\gamma_2} \right]^{-1}e(z,L_{\rm X})
\end{equation}
with evolution term 
\begin{eqnarray}
e(z,L_{\rm X}) = 
\begin{cases}
(1+z)^{p_1} & z<z_c(L_{\rm X}), \\
[1+z_c(L_{\rm X})]^{p_1-p_2}(1+z)^{p_2} & z\geq z_c(L_{\rm X}),
\end{cases}
\end{eqnarray}
and redshift cut-off
\begin{eqnarray}
z_c(L_{\rm X})  = 
\begin{cases}
z_c^* & L_{\rm X}\geq L_a, \\
z_c^* \left( \frac{L_{\rm X}}{L_a}\right)^\alpha & L_{\rm X}< L_a.
\end{cases}
\end{eqnarray}
The parameter values are:
$A = 5.04\times 10^{-6}$~Mpc$^{-3}$;
$L_a = 10^{44.6}$~erg~s$^{-1}$;
$L_* = 10^{43.94}$~erg~s$^{-1}$;
$\gamma_1 = 0.86$;
$\gamma_2 = 2.23$;
$p_1 = 4.23$;
$p_2 -1.5$;
$z_c^* = 1.9$ and
$\alpha = 0.335$.

In calculating the XLF from a given evolving BHMF, we will have to correct for the sampling bias of obscured, Compton-thick AGN, which is not included in the XLF of \cite{2003ApJ...598..886U}. The contribution of obscured sources is taken into account by dividing the BHMF by $1.6$ before we use it to calculate the XLF. This correction is similar to the approach taken in \cite{2009MNRAS.396.1217R} and \cite{2008MNRAS.390..561C}, and assumes the obscured source contribution to the XLF is independent of luminosity, although the picture is likely to be more complicated and this simplifying assumption may be a reason for discrepancies we may find in the model and observations. \cite{2010Sci...328..600T} find the ratio of obscured to unobscured quasars to be $\sim 1$ and increasing at redshifts $z>1.5$. 

To calculate the XLF, we use the Eddington-ratio dependent bolometric correction $L_{\rm X} = L_{\rm bol}/\kappa(\lambda)$ based on \cite{2007MNRAS.381.1235V}. We set:
\begin{equation}
\kappa(\lambda) = \begin{cases}
19.3 & 0.1\leq \lambda, \\
36.9 & 0.1 < \lambda < 0.3,  \\
54.5 & 0.3\leq \lambda.
\end{cases}
\end{equation}

Assigning values for the free parameters $\eta$, $\tau_1$, $\tau_2$, and $M_{\rm{const}}$ allows us to calculate $\langle \dot{M}(M,t)\rangle$ and hence $\langle \dot{m}(M,t)\rangle$. However, we will need to assume a distribution for $\dot{m}(M,t)$ in order to calculate the XLF. The spread in $\dot{m}(M,t)$ is expected to be large for the following reasons. A significant fraction of BHs are known to be not accreting (or barely accreting) at a given time while others may be accreting highly. Also, the Eddington ratio $\lambda$ is a function of $\dot{m}$ and a wide spread in Eddington ratios has been observed [e.g. \cite{2006ApJ...648..128K} find a log-normal distribution of $\lambda$ for a fixed mass with a dispersion of $\sim0.3$~dex in the AGN and Galaxy Evolution Survey (AGES), and \cite{2010MNRAS.408.1714R} find a large number of low Eddington ratios $10^{-4} < \lambda < 10^{-1}$ in the \textit{Chandra Deep Fields}]. It appears that the more sensitive an observation is, the lower Eddington ratios may be found. The true spread in  $\dot{m}$, or equivalently, $\lambda$, may even be larger than those found by \cite{2006ApJ...648..128K} or \cite{2010MNRAS.408.1714R}, since here we need to include all galaxies with supermassive BHs in the distribution, even ones that are not accreting. In this analysis, we will approximate BHs that are not accreting with BHs that have negligible accretion $\dot{m}$ (their contribution to the total kinetic feedback will also be small).

We assume a log-normal probability distribution for $\dot{m}(M,t)$. The distribution is centred around $\langle \dot{m}(M,t) \rangle-\frac{1}{2}\sigma^2\ln 10$ with standard deviation $\sigma$ (the centre is chosen so that the mean of the distribution is $\langle \dot{m}(M,t) \rangle$, which is what we want). The standard deviation $\sigma$ is the final free-parameter in our model. The log-normal distribution is a reasonable assumption because Eddington ratios appear to be roughly distributed in a log-normal fashion \citep{2006ApJ...648..128K,2010MNRAS.408.1714R} so the $\dot{m}$ distribution should be similar as well when we convert it using the relation between $\dot{m}$ and $\lambda$. The shape of the left-tail of the distribution may not be well constrained by observations, but it has a small effect on the average of the distribution and objects in the left wing do not end up contributing to the $10^{41.5}$--$10^{46.5}$~erg~s$^{-1}$ XLF. Knowing the probability distribution $D(\dot{m}(M,t))$ of $\dot{m}(M,t)$, one can obtain the Eddington ratio probability distribution using $D(\dot{m}(M,t)) d\dot{m} = D(\lambda(M,t)) d\lambda$. To obtain the distribution of $L_{\rm X}$, we separate the distribution $D(\lambda(M,t))$ into three distributions: $0.1\leq \lambda$, $0.1 < \lambda < 0.3$, $0.3\leq \lambda$ (because of the discontinuities in the bolometric correction factor $\kappa(\lambda)$), convert each using $D(L_{\rm X}(M,t)) dL_{\rm X}  = D(\lambda(M,t)) d\lambda$, and sum. Knowing the probability distribution of $L_{\rm X}$ for given $M$ and $t$, as well as the number density of sources from the evolving BHMF at time $t$ with masses between $M$ and $M+\Delta M$ provides the information needed to construct the XLF.

Calculating an XLF with the method described above for various values of the free-parameters, we find that $\sigma$ does indeed have to be large ($>0.3$~dex at least) in order to construct reasonable XLFs (the best-fitting value for all parameters is reported in \S~\ref{sec:results}). Simply put, $\sigma \lesssim 0.3$~dex does not produce objects with a wide enough distribution of $L_{\rm X}$ to cover at least $10^{41.5}$ to $10^{46.5}$~erg~s$^{-1}$, which is the domain of the observationally derived XLF. The number density of the BHs derived from an evolving BHMF is $1$--$2$ orders of magnitude higher than the number density of BHs in the observed XLF (for example, at $z=0$, the number density of BHs is $2\times 10^{-2}$~Mpc$^{-3}$ and the number density of X-ray sources with $10^{41.5}\leq  L_{\rm X}($erg$~$s$^{-1})^{-1} \leq 10^{46.5}$ is $3\times 10^{-4}$~Mpc$^{-3}$). A large fraction of galaxies with BHs in fact do not contribute to the observed $10^{41.5}$--$10^{46.5}$~erg~s$^{-1}$ XLF, which is accounted for by having a large spread as dictated by $\sigma$ so that a considerable fraction of sources are accreting well below the critical accretion rate $\dot{m}_{\rm crit}$.

Numerically calculating the XLF as a sum of Gaussians for given values of the free parameters $\eta$, $\tau_1$, $\tau_2$, $M_{\rm{const}}$, and $\sigma$, we find a good fit for the free parameters by minimizing the sum of the squared differences of 
$\log \frac{\rmn{d} \phi_{\rm observed} (L_{\rm X},z)}{\rmn{d} \log L_{\rm X}}$ and
$\log \frac{\rmn{d} \phi_{\rm derived} (L_{\rm X},z)}{\rmn{d} \log L_{\rm X}}$
at $\log (L_{\rm X} ($erg$~$s$^{-1})^{-1}) = 41.5, 42.0, 42.5, \ldots, 46.5$. The final model is to be presented in \S~\ref{sec:results}.

\subsection{Kinetic Feedback}

Once having found parameters for an evolutionary scenario that is able to reproduce observations (XLF and the local BHMF), the total amount of volume-averaged kinetic feedback may be calculated as a function of redshift, as well as a KLF, if we make an estimate for the fraction, $f$, of high accretion rate sources in HK mode (as opposed to HR mode). This fraction $f$ is the fraction of high-accreting sources that are radio-loud at a given cosmic time. We will present two calculations, one with a fiducial value of $f=0.1$ across all redshifts and another with a refined estimate of $f$ as a function of redshift needed to match the observed RLF for powerful, jetted radio sources.

The total kinetic energy per unit comoving volume at time $t$ is calculated as:
\begin{equation}
L_{\rm{kin,tot}}(t) = \int  \langle \lambda_{\rm kin}(M,t)\rangle L_{\rm Edd}(M) \times \frac{\partial N(M,t)}{\partial M} \,dM
\end{equation}
where
\begin{equation}
\langle \lambda_{\rm kin}(M,t)\rangle  = \int   \lambda_{\rm kin}(\dot{m})  D(\dot{m}(M,t))\,d\dot{m}.
\end{equation}
The total kinetic energy per unit comoving volume may be broken down into the HK, HR, and LK contributions as well.

The KLF is calculated similarly to the way it is done for the XLF. The distribution of $\dot{m}(M,t)$ can be converted to a distribution for $\lambda_{\rm kin}(M,t)$ for both expressions for $\lambda_{\rm kin}(M,t)$ (due to the two high-accretion modes). The KLF may be calculated for both distributions and a weighted average may be taken according to the assumed value for $f$ to obtain the total KLF.

\subsection{FR II sources and feedback}

From the calculated KLF, we can estimate the kinetic feedback due to powerful classical double-lobed FR~II \citep{1974MNRAS.167P..31F} sources and the amount of energy that goes into doing $PdV$ work in expanding the lobes. Each lobe is powered by a jet with power $Q_{\rm jet} = L_{\rm kin}/2$. We will consider sources with $Q_{\rm jet}>5\times 10^{37}$~W to exhibit FR~II jets, which is the lower end of the distribution of jet power found in \cite{1999AJ....117..677B} for analytic models of FR~II radio lobes to match observations. We can integrate the KLF (multiplied by $L_{\rm kin}$) above luminosities $L_{\rm kin}=10^{38}$~W to find the total power going into FR~II jets as a function of redshift. We estimate the fraction of the jet power that goes into $PdV$ work to expand the lobes from the analytic model developed for FR~II sources in \cite{2011MNRAS.413.1107M}. The fraction is found by calculating $\int_0^{t_{\rm j}} p_{l} \,dV_{\rm l} / (Q_{\rm jet} t_{\rm j})$ for a typical jet lifetime of $t_{\rm j}=5\times 10^8$~yr and environmental parameters $\beta=1.5$, $a_0=10~{\rm kpc}$ and $\rho_0=1.67\times 10^{-23}~{\rm kg}~{\rm m}^{-3}$ \citep{1999AJ....117..677B} and injection parameters $\gamma_{\rm min} = 1$, $\gamma_{\rm max} = 10^6$ and $p=2.14$ \citep{2011MNRAS.413.1107M}. We find that the fraction, $F$, of power that goes into the expansion of the lobes for this set of parameters, as a function of jet power $Q_{\rm jet}$ in units of W, can approximately be parametrized as:
\begin{equation}
F(Q_{\rm jet}) = 2.4\cdot 10^7 Q^{-0.21}.
\end{equation}
For a typical jet power of $Q_{\rm jet} = 10^{38}$~W, the fraction is approximately $1/6$. This value is reasonable compared to the calculation in \cite{1997MNRAS.292..723K} (see their equation ($17$)) based on a simpler model of self-similar evolution of lobes. In our non-self-similar analytic model for evolution of FR~II lobes, this fraction does depend somewhat on jet power and time along the evolution, and this dependence is being investigated in another paper (Mocz et al., in preparation). The remainder of the power that is supplied by the jet (other than expansion work by the lobes) goes into internal energy stored in the lobes, internal energy in the hotspot (which is negligible), expansion work done by the hotspot, and energy lost due to synchrotron radiation in the radio and up-scattering of cosmic microwave background (CMB) photons to X-ray energies via inverse-Compton (IC) scattering. We show in \cite{2011MNRAS.413.1107M}, that after the jets switch off, the radio emission due to synchrotron radiation plummets rapidly but the lobe may still be overpressured and will continue to expand, although not as rapidly as if the jet were still on. This means that there will still be $PdV$ work done after the lobes are not detectable in the radio. Using the analytic model developed in  \cite{2011MNRAS.413.1107M}, we find that if a jet is turned on at $z=2$ for $t_{\rm j}=5\times 10^8$~yr, the ratio of $PdV$ work the lobes do after the jets switch off until the present epoch $z=0$ is a considerable fraction, $\sim 0.6$, of the work done during the time the jet was on. The fraction of the total jet energy $Q_{\rm jet} t_{\rm j}$ lost due to synchrotron radiation is found to be negligible, $\sim 0.002$. The energy lost to inverse-Compton scattering of the CMB (ICCMB) may be more significant though still small, for example $\sim 0.03$ at $z=2$, but it is important to note that this fraction depends on redshift due to the $(1+z)^4$ dependence in the CMB energy density as a function of redshift.

\subsection{RLF for FR II sources}

In addition, with a KLF and a model for FR~II radio sources, as well as an assumption about the jet duty cycles, we can derive an RLF from our population and compare with the observed RLF of \cite{2001MNRAS.322..536W}. Comparison of the derived and observed RLF then tells us how to refine the jet duty cycle assumption $f$ as a function of redshift. 

We assume that for a given redshift, the radio emission from an FR~II source of jet power $Q_{\rm jet}$ is randomly chosen between $t=0$ and $10t_{\rm j}$ in the evolution of the source (the radio emission plummets very shortly after $t=t_{\rm j}$). The time $t=10t_{\rm j}$ corresponds to a time when the pressure in the lobes become comparable to the surrounding intergalactic medium (IGM; the lobes do remain overpressured for some time after jets switch off; \citealt{2011MNRAS.413.1107M}). This is synonymous with saying that $f=0.1$ of sources with potentially powerful jets have their jets turned on at a given time. The radio emission is calculated using the model in \cite{2011MNRAS.413.1107M}. The RLF can then be derived from the KLF which tells us the number of sources with jet powers between $Q_{\rm jet}$ and $Q_{\rm jet}+\Delta Q_{\rm jet}$ above $Q_{\rm jet}>5\times 10^{37}$~W (the minimum FR~II jet power).

The calculated KLF can be compared with the observed one and used to constrain $f$ as a function of $z$. We can then eliminate our initial $f=0.1$ assumption and derive more accurate an estimate for FR~II feedback.

\section{Results}\label{sec:results}

\begin{figure}
\centering
\includegraphics[width=0.47\textwidth]{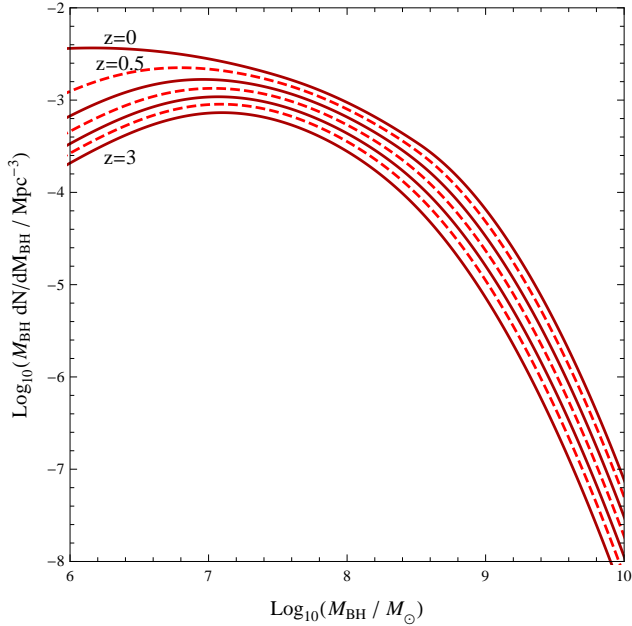}
\caption{The evolving BHMF that best reproduces the observed XLF. Anti hierarchical growth is present in the evolution. The alternating solid/dashed lines show the evolving BHMF from $z=0$ to $z=3$ in increments of $\Delta z=0.5$.}
\label{fig:bhmf}
\end{figure}

\begin{figure}
\centering
\includegraphics[width=0.47\textwidth]{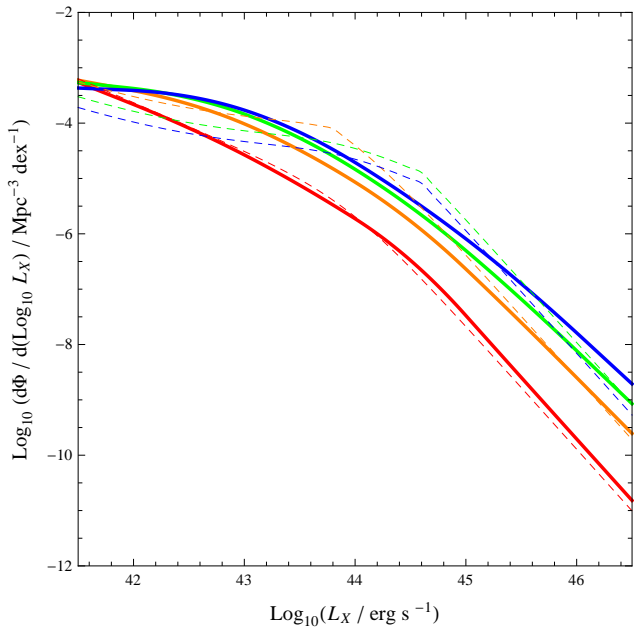}
\caption{Comparison of observed (dashed) and derived (solid) XLFs at $z=0$ (red), $1$ (orange), $2$ (green) and $3$ (blue). The two are in relatively good agreement. Our simple assumption of a Gaussian spread in $\dot{m}$ does not produce quite as sharp a `knee' as in the observed XLF. Our simple model slightly over-predicts the XLF at $z=3$.}
\label{fig:xlf}
\end{figure}

\begin{figure}
\centering
\includegraphics[width=0.47\textwidth]{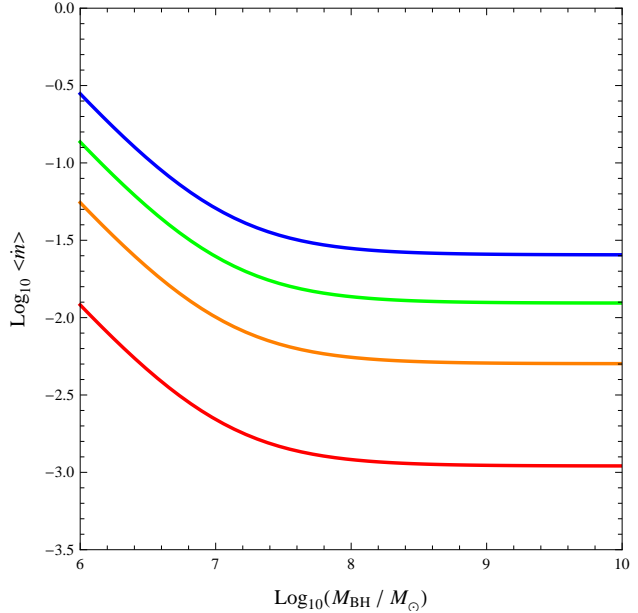}
\caption{The average value of $\dot{m}$ in our model at $z=0$ (red), $1$ (orange), $2$ (green) and $3$ (blue) as a function of BH mass, $M$. High and low mass BHs evolve differently (note the change in slope, due to the inclusion of $M_{\rm const}$ in our evolution scenario) which allows for anti hierarchical growth. The parameters $\tau_1$ and $\tau_2$ (which are found to be roughly the same) determine how quickly the low and high ends of the plot of $\dot{m}$ changes with time.}
\label{fig:mdot}
\end{figure}

We carry out the steps mentioned in \S~\ref{sec:evo} to evolve the BH population backwards in time and find that we can reasonably reproduce the observed XLF with our simple model. We find best-fitting parameters 
$\eta=0.08\pm 0.02$,
$\tau_1=10^{9.9\pm0.1}$~yr, 
$\tau_2=10^{9.9\pm0.1}$~yr, 
$M_{\rm{const}} = 10^{7.0\pm1.0}~M_\odot$, and
$\sigma=0.9\pm0.1$~dex, with $1$-$\sigma$ errors reported.
The evolution of the BHMF is shown in Figure~\ref{fig:bhmf}. A non-zero value for $M_{\rm{const}}$ is favoured, which leads to anti hierarchical growth, but the two time-scales $\tau_1$ and $\tau_2$ converge. The derived XLF from the population of BHs is presented in Figure~\ref{fig:xlf}, which we see agrees well with the observed XLF considering the simplifying assumptions we have made. The largest discrepancies (factor of $3-10$) between the predicted and observed XLF occur at the `knee' feature of the observed XLF which is not possible to produce with the simple assumption of a smooth log-normal distribution of $\dot{m}$ and the low-luminosity end of the XLF, but this is generally within the $1$-$\sigma$ Poisson errors in the observational determination of the XLF. A significant spread in the log-normal distribution of $\dot{m}$, namely $\sigma=0.9\pm0.1$~dex, is required to match observations well. The average values of $\dot{m}$ as a function of BH mass at various redshifts are plotted in Figure~\ref{fig:mdot}. The average values of $\dot{m}$ are well below $1$ for these redshifts.

\begin{figure}
\centering
\includegraphics[width=0.47\textwidth]{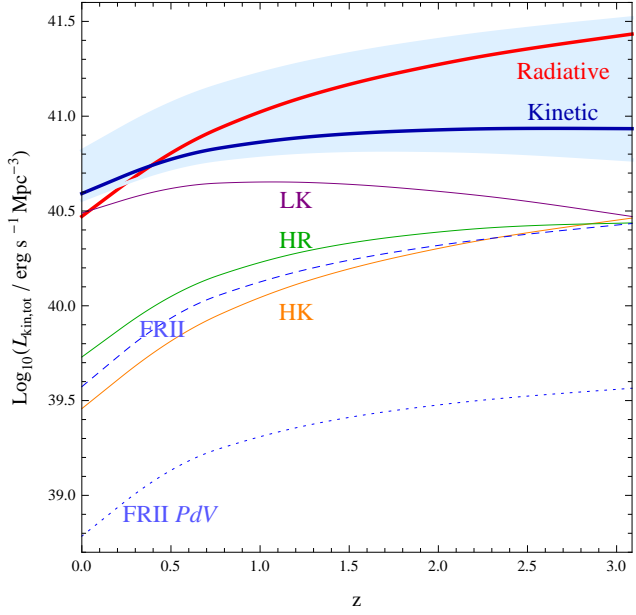}
\caption{The energies of various components of BH feedback per comoving volume as a function of redshift. Contribution due to only unobscured sources is shown; to estimate the effect of the obscured sources, multiply by $1.6$. The total radiative feedback (total bolometric luminosity of sources) is indicated by the thick red line. The light blue shaded region shows the possible total kinetic feedback if one varies the fraction $f$ of sources accreting above $\dot{m}_{\rm crit}$ in HK mode from $f=0$ to $1$. The thick dashed blue line shows the kinetic output for $f=0.1$. The orange, green, and purple lines show the breakdown contribution of LK, HK and HR sources, respectively, to the total kinetic feedback. The kinetic feedback due to FR~II sources ($L_{\rm kin} = 2Q_{\rm jet}>10^{38}$~erg~s$^{-1}$) is shown in dashed blue and closely follows the HK line. The fraction of the FR~II kinetic power that goes into $PdV$ work to expand the lobes is shown in dotted blue.}
\label{fig:KE}
\end{figure}

\begin{figure}
\centering
\includegraphics[width=0.47\textwidth]{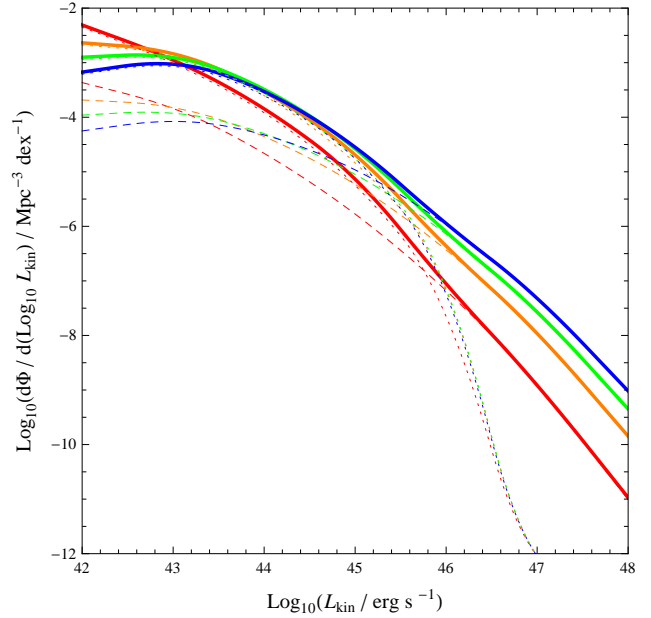}
\caption{The derived KLF from our simple evolutionary scenario, assuming a fraction $f=0.1$ of highly accreting sources are HK and a fraction of $1-f=0.9$ highly accreting sources are HR. The KLF is shown at $z=0$ (red), $1$ (orange), $2$ (green) and $3$ (blue). The dashed lines correspond to the HK plus $f=0.1$ of LK sources and the dotted lines correspond to the HR plus $1-f=0.9$ of LK sources. Sources with kinetic luminosities $L_{\rm kin}>10^{45}$~erg~s$^{-1}$ (corresponding to the minimum jet energy we consider to create FR~II type lobes) are mostly HK sources.}
\label{fig:KLF}
\end{figure}

\begin{figure}
\centering
\includegraphics[width=0.47\textwidth]{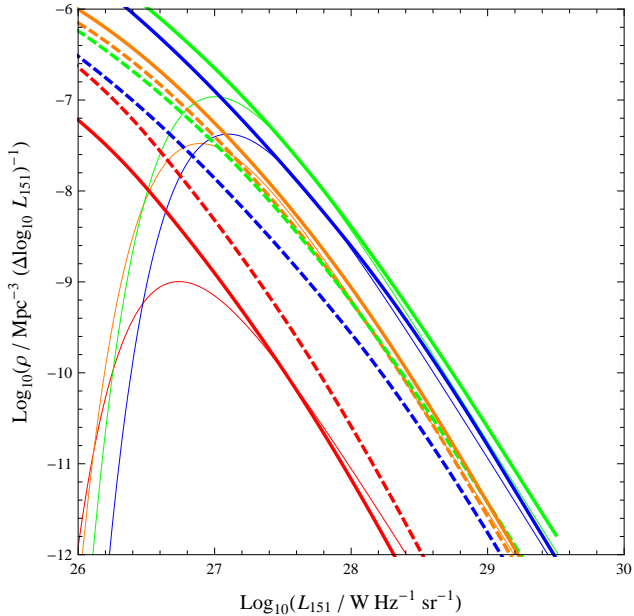}
\caption{The derived RLF for FR~IIs from the KLF. The thick dashed lines correspond to the derived RLFs at $z=0$ (red), $z=1$ (orange), $z=2$ (green) and $z=3$ (blue) assuming that $f=0.1$ of such sources are switched on at all redshifts. The thin lines are the observed RLFs for the population of the brightest FR~IIs from \protect\cite{2001MNRAS.322..536W}. There is discrepancy between the two sets of RLFs, due to the $f=0.1$ assumption. Underestimation means that $>0.1$ potential FR~II sources are switched on at that redshift. Overestimation means that $<0.1$ sources are on. The RLF is corrected by varying $f$ with redshift and the corrected RLF is shown in thick solid lines.}
\label{fig:RLF}
\end{figure}

\begin{figure}
\centering
\includegraphics[width=0.47\textwidth]{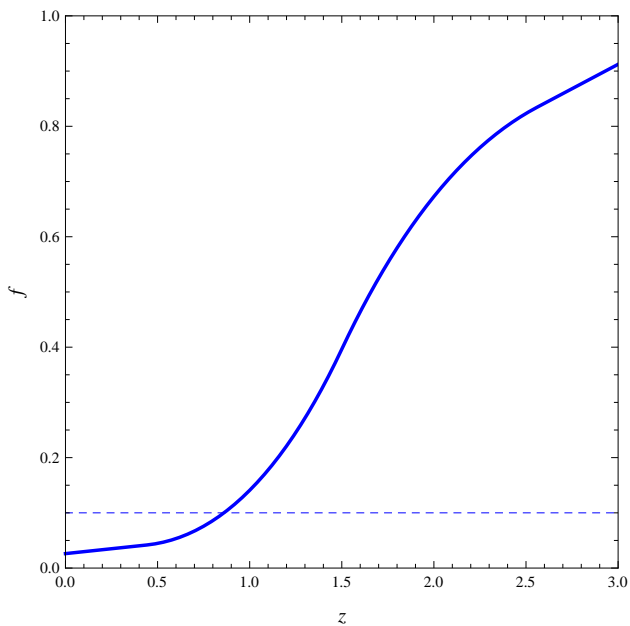}
\caption{Approximate fraction of powerful radio sources switched on (jet duty cycle) as a function of redshift so that our derived RLF matches the observed one. Dashed line shows a fiducial $0.1$ value for comparison. This amounts to correcting the solid kinetic line ($f=0.1$) in Fig.~\ref{fig:KE} to the observed (dot-dashed) kinetic.}
\label{fig:F}
\end{figure}

\begin{figure}
\centering
\includegraphics[width=0.47\textwidth]{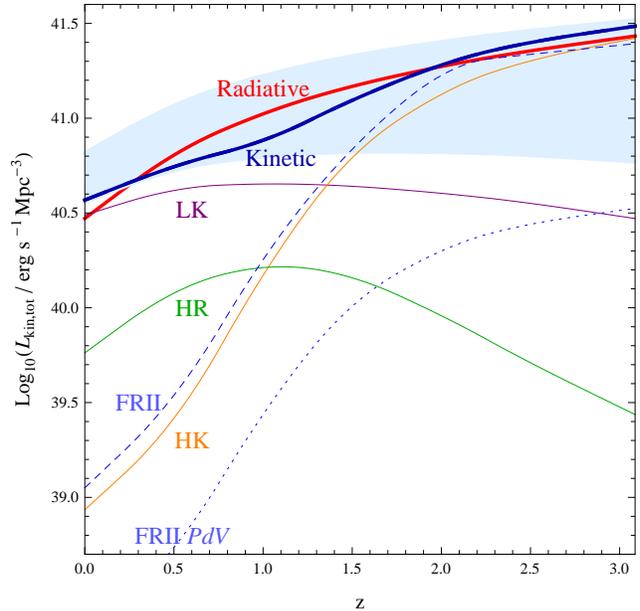}
\caption{Same as Figure~\ref{fig:KE} except with the refined assumption that $f$ can vary with redshift, which is estimated from the observed RLF.
It is interesting to note that the total kinetic output follows the radiative energy output closely. Also, with this refined estimate for $f$ we see that HK mode feedback becomes much more dominant at $z>1.5$.}
\label{fig:KE2}
\end{figure}

\begin{figure}
\centering
\includegraphics[width=0.47\textwidth]{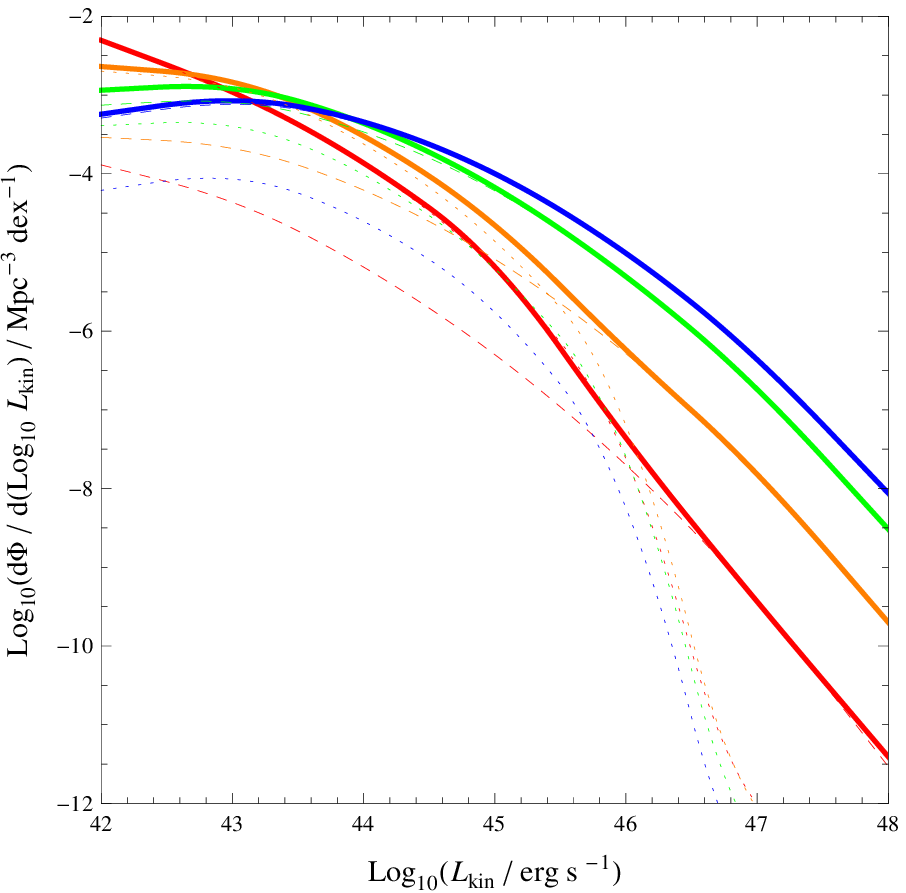}
\caption{Same as Figure~\ref{fig:KLF} except with the refined assumption that $f$ can vary with redshift, which is estimated from the observed RLF.}
\label{fig:KLF2}
\end{figure}

Figure~\ref{fig:KE} shows the derived total kinetic feedback (and components due to different modes of accretion, and FR~II sources) per unit comoving volume as a function of redshift assuming $f=0.1$ [meaning $1/10$ of high-accretion sources have jets (i.e., are in HK mode) at a given cosmic time]. Figure~\ref{fig:KLF} shows the accompanying derived KLF. These two figures are refined by then estimating an evolving duty cycle $f=f(z)$ from deriving the RLF for FR~II sources and comparing it to the observed RLF (Figure~\ref{fig:RLF}). The evolution of the total kinetic feedback with this better informed duty cycle (plotted in Figure~\ref{fig:F}) is presented in Figure~\ref{fig:KE2} and the corresponding KLF is shown in Figure~\ref{fig:KLF2}. These are the three main figures of our paper.

Our calculation of the RLF for FR~II sources matches the observed RLF at $z=1$ under the simple assumption that a fraction $f=0.1$ of powerful radio sources are switched on at a given time, corresponding to the picture that all sources have episodic jet activity with jets switched on for $5\times 10^8$~yrs and enough time between events so that the lobes stop being overpressured. With the $f=0.1$ assumption we overestimate the RLF at the present epoch, which means that a large fraction of once-powerful radio sources are switched off today, perhaps permanently if the fuel supply has depleted and nothing triggers more gas infall. At $z>1$, we underestimate the RLF, meaning that many radio sources have had to be switching on at redshifts $1.5<z<3$ (the quasar era; this corresponds to a length of time of $4t_{\rm j}$) and that a large fraction ($>0.1$, close to $1$ in fact) of high-accretion sources that could host powerful jets were in fact active. Figure~\ref{fig:F} shows, as a function of redshift, the fraction of high-accretion sources that need to be turned on for the derived and observed RLFs to match. The fraction varies from $>0.8$ at $z=3$ to $0.02$ at $z=0$. A revised estimate of the total kinetic feedback is also shown in Figure~\ref{fig:KE2}. In such case, the amount of kinetic feedback rises to approximately match the radiative feedback. The total kinetic feedback is then dominated by FR~II sources at redshifts above $z\sim 1.5$.

\section{Discussion}\label{sec:disc}

We are able to reconstruct the observed XLF with our simple model for the growth of BHs. We integrate the local BHMF backwards according to a simple evolutionary scheme to obtain the evolving BHMF and average accretion rate as a function of BH mass and time. In our model, anti hierarchical BH growth is present due to a same quantity of mass, $M_{\rm const}$, being accreted by all BHs irrespective of BH mass. The term $M_{\rm const}$ may be interpreted as a uniform source of matter available to be accreted by all BHs. Possibly such matter may be due to expelled material from an earlier, more violent, merger-dominated accretion history of BHs before redshift $z_{\rm i}$.

The evolutionary scenario we described in this paper is simple with a limited number of free-parameters, which is not expected to fully describe in detail the evolutionary histories and feedback of individual supermassive BHs. However, the free-parameters all have distinct effects in the model, which avoids degeneracies and provides intuition on how the BHMF and XLF evolves. The free parameters that describe the accretion history of supermassive BHs in our model are $M_{\rm i}$, $\tau_1$, $\tau_2$, and $\eta$. The evolution of BHs in the high-mass end ($M>M_{\rm i}$) is only primarily determined by $\eta$ which adjusts how much matter is accreting on to the BHs and $\tau_1$ which describes how this amount changes with time. BHs in the low-mass-end ($M<M_{\rm i}$) are allowed to evolve differently (different rates and change in rates from the high-mass counterparts) with the inclusion of the other two parameters $M_{\rm i}$ and $\tau_2$ to allow for the possibility of anti hierarchical growth. Finally, we chose a log-normal distribution of $\dot{m}$ for a given mass characterized by a single free-parameter, the standard deviation $\sigma$, to minimize degeneracies when matching the predicted and observed shape of the XLF. Supposing we have two different evolving BHMFs, it becomes impossible with a single parameter $\sigma$ to produce two XLFs with the same slopes {\it and} the same integral constraints. A more sophisticated model may improve the match between the observed and predicted XLF, but would be prone to degeneracies in the best-fitting free-parameters. 

Knowing the true distribution of $\dot{m}$ for a given BH mass, which in general may change its shape and spread with redshift, is the most difficult to directly determine from the constraints we impose. We chose to investigate a simple log-normal distribution to understand what spread in the distribution the observations suggest. However, a refinement of the distribution would improve the predicted shape of the XLF function (although introduction of more than a single free parameter would introduce degeneracies in the fitting of the XLF shape and hence larger uncertainties). A modified distribution for $\dot{m}$, such as a log-normal with a power-law tail for low $\dot{m}$, as suggested from observations by \cite{2009MNRAS.397..135K}, could better reproduce the `knee' feature in the XLF and the XLF at the lower luminosities (generally corresponding to lower-mass and smaller $\dot{m}$ BHs). Such a modified distribution would mostly affect the number density and distribution of low-$\dot{m}$ sources, hence not affecting much our estimate of jet duty cycle with redshift, which is determined by the evolution of high-$\dot{m}$ sources.

The large spread we find in $\dot{m}$, namely a standard deviation of $\sigma=0.9\pm0.1$~dex, helps account for BHs that are not or negligibly accreting. This spread corresponds to a $\sim 0.9$~dex standard deviation spread in Eddington ratios. A significant fraction of sources thus have $\lambda<10^{-3}$, corresponding to sources with little/negligible accretion activity and sources that do not contribute to the XLF. As the average accretion rate, $\langle \dot{m}(M,t)\rangle$, decreases with time in our model, a higher fraction of sources have little activity ($\lambda<10^{-3}$) as we reach $z=0$.

From our evolution scenario we can quantify the amounts of different modes of feedback and learn which one dominates. The amount of kinetic feedback may exceed radiative feedback in two cases. First, if the fraction of HK sources is high, then the kinetic feedback will always exceed radiative feedback at all redshifts (if we assume the BH accretion physical model that is presented Figure~\ref{fig:scheme}). Secondly, once most of the sources are accreting in LK mode, kinetic feedback will exceed radiative feedback, which occurs in our calculations for $z<0.4$. In our calculations, with the refined assumption of the duty cycle $f$ changing throughout cosmic time, the total kinetic and radiative feedback closely match each other, which is somewhat of a coincidence caused by both the average accretion rates and duty cycle $f$ decreasing with smaller redshift.

The total kinetic feedback may be decomposed into HK, HR, and LK mode feedback. The decomposition is important to understand because different fractions of the energy from the three feedback modes may go into actually heating the interstellar medium (ISM) and IGM. We see from Figure~\ref{fig:KE2} that, with our estimate for evolving jet duty cycle, much of the total kinetic output energy  at $z<1.5$ comes from the LK mode, and its contribution to the total power may increase up to $z\sim 0.8$ even though the total kinetic feedback decreases with time. Our calculations highlight the importance that sources in LK mode may have in feedback. Since the total feedback in LK mode is higher than HK and HR modes at $z<1.5$, even if a smaller fraction of the LK mode kinetic energy goes into heating the gas in the host galaxy it could still be more important. Sources accreting in LK mode would correspond to FR I radio sources. \cite{2010ApJ...710..743D} argued, using comparisons with observed data, that feedback in FR I sources may occur due to AGN outflows that are strongly decelerated and become fully turbulent sonic or subsonic flows due to their interaction with the surrounding medium.

Our calculations also show, however, that at earlier redshifts $z>1.5$, when the jet duty cycle is large ($>0.5$), the total volume-averaged energy feedback from the HK mode (classical double-lobed FR~II type feedback) is dominant. These sources expand $10$--$1000$ kpc lobes and do $PdV$ work on the surrounding IGM. The amount of $PdV$ work during the time the jet is on is found to be roughly $1/6$ of the kinetic energy that goes into the jets. It is unclear how much of this energy goes into heating the central galaxy, which can be presumably small since the hotspots and lobes are far from the galaxy for the majority of the time the jet is switched on, and rather these giant lobes give considerable feedback to the IGM. Galaxies that show FR~II processes must have had feedback heating their ISM as well, and perhaps much of this feedback comes either from radiative feedback or the kinetic feedback during the time spent in HR mode. If the HR mode is the predominant form of feedback in the ISM for FR~II type sources, then a much larger fraction of the output energy is needed to heat the ISM than would be needed if HK mode were the culprit, as the total volume-averaged energy from HR feedback is an order of magnitude below HK feedback at $z>1.5$.

With our approach to determine the total kinetic feedback (which uses the RLF along with a model for FR~II evolution to estimate the jet-duty cycle as a function of redshift) we find a significantly higher estimate than previous studies by almost an order of magnitude at $z>1.5$. We predict that the total volume averaged kinetic feedback closely traces the radiative feedback rather than being about an order of magnitude below it. \cite{2009MNRAS.395..518C} also calculate radiative and kinetic feedback as a function of redshift, using the bolometric LF of \cite{2007ApJ...654..731H} to find radiative feedback and the RLF of \cite{1990MNRAS.247...19D} for extragalactic radio sources to estimate mechanical power in kinetic feedback. Their calculated ratio for radiative to kinetic feedback agrees with the earlier work of \cite{2008MNRAS.383..277K}, who also use an RLF directly for calculating the mechanical feedback (unlike our work). We find a higher result for total kinetic feedback because some LK sources may be missing from the RLF, as well as a fraction of bright sources due to the `youth-redshift' degeneracy described in \cite{1999Natur.399..330B}, which says that older radio sources, as they expand, decrease their synchrotron radiation in the radio (due to lowering of the magnetic field strengths during expansion) and fall below the flux limit of the surveys used to construct the RLF and thus go undetected. A model for the time evolution of the radio emission of radio sources is needed to accurately have a handle on the `youth-redshift' degeneracy, which is part of our approach. Our calculation of radiative feedback, unlike total kinetic feedback, agrees very well with the calculation in \cite{2009MNRAS.395..518C}. The amount of $PdV$ work by FR~II sources is comparable to the amount of cavity work calculated in \cite{2009MNRAS.395..518C} as a function of redshift. But we find that the total kinetic feedback is significantly higher than the fraction that goes into expanding lobes, and is comparable to the total radiative feedback.

Kinetic radio-mode feedback is previously overlooked and potentially very important for the growth and feedback in galaxies on a cosmological scale. The review by \cite{2012ARA&A..50..455F} identifies FR~II feedback (powerful, jetted radio outbursts) as an ill-understood feedback mode which can deposit the kind of energy that can disrupt even a group of galaxies. In comparison to FR~I kinetic feedback and radiative (quasar) feedback, FR~II feedback has been little discussed in the literature. Here we have provided a first estimate of the amplitude of FR~II feedback. From Figure~\ref{fig:KE2} we see that its amplitude may be very significant above $z\sim1.5$, rivalling radiative output, much of which passes beyond the host galaxy. Accounting for this level of FR~II type feedback in cosmological simulations may refine our picture of galaxy evolution.

\section*{Acknowledgements}
KMB and ACF thank the Royal Society for support.

\bibliography{mybib}{}

\bsp
\label{lastpage}
\end{document}